\documentclass[a4paper,14pt]{article}
\usepackage{graphicx}
\usepackage{amsmath}
\title{Analysis of $ B^{+}_{c}\rightarrow J/\psi a_{1}(1260)^{+}$ in pertubative QCD approach}
\author{Fatemeh Najafi\footnote{fatemeh.najafi@students.semnan.ac.ir},
 Hossein Mehraban\footnote{hmehraban@semnan.ac.ir}\\
Physics Department, Semnan University\\
P.O.Box 35195-363, Semnan, Iran}
\begin{document}
\maketitle
\begin{abstract}
In this paper, we analyse the hadronic decay of $ B^{+}_{c}\rightarrow J/\psi\ a_{1}(1260)^{+}$ in pertubative QCD approach (pQCD), where $ a_{1}(1260) $ is a axial-vector meson and $ J/\psi $ is a vector meson. The experimental data of the branching ratio of this decay is less than ($ 1.2\times 10^{-3} $). 
We obtained that the branching ratio is $ (1.02^{+0.04+0.03+0.01}_{-0.08-0.05-0.01})\times 10^{-3} $
and 
has a good
agreement with the experimental result.
\end{abstract}

\section{Introduction} 
The $B_{c} $ meson is a ground state of $ \bar{b}c $ system and 
is the only heavy meson embracing two heavy quarks b and c. Because lifetime, mass and the relative Cabibbo-Kobayashi-Maskawa (CKM) matrix element between b and c quark are diffrent, the decay rate of the two quarks is different. 
Therefore studying of $ B_{c} $ decays provide information about CP violation and standard model.
Since the $B_{c} $ meson carries explicit
flavor, it can not annihilate via strong interaction or electromagnetic interaction like the
mesons consisting of $ c\bar{c} $ or $ b\bar{b} $. It can only decay via weak interaction that is an
ideal platform to study weak decays of heavy quarks.
Compared with the $ B_{u,d,s} $ mesons, the decays of the $ B_{c}
$ meson are rather different from those of $ B$ or $ B_{s} $ meson, since
in $ B_{c} $ meson both b and c can decay while the other serves
as a spectator, or annihilating into pairs of leptons or light
mesons. Exclusive modes containing of $ B_{c}\rightarrow
h_{1}h_{2} $ decays ($ h_{i} $ are the vector, axial-vector,pseudo-scalar, tensor and scalar). Two-body hadronic $B_{c}$ decay meson has 
studied by many author~\cite{1,2,e,t,f,g,h}.
In this research, the two
body hadronic decay $ B^{+}_{c}\rightarrow J/\psi \
a_{1}(1260)^{+}$ are studied, where $ a_{1}(1260) $ and $ J/\psi $ are
axial-vector and vector respectivity. Axial-vector mesons are
mesons with the quantum numbers $ J_{P}=1^{+} $. In the quark
model, there are two distinct types of axial-vector mesons,
namely, $^{3}P_{1}$ and $^{1}P_{1}$,which carry the quantum
numbers $ J^{PC}=1^{++} $ and $ J^{PC}=1^{+-} $ respectively.
Experimentally, the $ J^{PC}=1^{++} $ nonet consists of $
a_{1}(1260) $, $ f_{1}(1285) $, $ f_{1}(1420) $, and $ K_{1A}$,
while the $ J^{PC}=1^{+-}$ nonet has $ b_{1}(1235) $, $
h_{1}(1170) $, $ h_{1}(1380) $ and $ K_{1B} $. In the $ SU(3) $
flavor limit, these mesons can not mix with each other. Due to the
$ G $-parity the chiral-even two-parton light-cone distribution
amplitudes of the $^{3}P_{1}$ ($^{1}P_{1}$) mesons are symmetric
(antisymmetric) under the interchange of the momentum fractions of
quark and anti-quark in the $ SU(3) $ limit. The $^{3}P_{1}$ meson
behaves in a similar way to the vector meson; this is not the case
for the $^{1}P_{1}$ meson. For the latter, its decay constant
vanishes in the $ SU(3) $ limit.
Since c quark’s mass is about one third of b quark and $ m_{c}/m_{B_{c}}\sim0.2 $, therefore b quark in a $ B_{c} $ carry the large part of the energy. This assumption allows us to employ the $ k_{T} $ factorization theorem to the b decay in $ B_{c} $ meson.
In this work we shall study
$ B^{+}_{c}\rightarrow J/\psi \ a_{1}(1260)^{+}$ decays in the perturbative QCD approach based on the $ k_{T} $ factorization\cite{sp}. 
We know that the light quark in B meson is soft, while it is collinear in the
final state light meson, so a hard gluon is necessary to kick the light spectator
quark in the B meson. By keeping the transverse momentum $ k_{T} $ in the quark
and gluon propagators, the end point singularity in the collinear factorization
can be eliminated. Because of the additional energy scale introduced by the
transverse momentum, double logarithms will appear in the QCD radiative corrections.
With the Sudakov resummation, we can include the leading double
logarithms for all loop diagrams, in association with the soft contribution. This
makes the pQCD approach more reliable and consistent.
The appealing feature of the
pQCD factorization is that form factors can be
computed in terms of wave functions and hard kernels. 
Although the $ k_{T} $
factorization is not gauge invariant and produces an 
IR finite hard kernel \cite{fan}, the pQCD method provides a good platform
to study of B meson two-body non-leptonic decays\cite{fan2}.
There are three scales in the B meson non-leptonic decays $ M_{W} $, $ m_{b} $ and $ 1/b $. 
The electroweak decay of B meson is characterized by the W boson mass $ M_{W} $. 
The
second scale $ m_{b} $ reflects the scale of energy in the decay.
Since the b quark decay scale $ m_{b} $
is much smaller than the electroweak scale $ M_{W} $, the QCD corrections are nonnegligible.
The third scale $ 1/b $ is the factorization scale, with 
$ b$ being the conjugate variable of parton transverse momenta.
Dynamics below
$ 1/b $ is regarded as being completely nonperturbative, and can be parametrized into a meson wave fucntion $\phi(x) $. The meson wave functions are not calculable in pQCD. But they are universal, channel independent. 
This paper is
organized as follows. In section II, we study decay constant of
axial-vector, form factor , wave function and distribution
amplitudes for $ B_{c} $, vector and axial-vector mesons. In
section III, we calculate the branching ratio and CP violation for this decay, section IV contain input quantitie parameters and
section V contain the main conclusion.
\section{Theorical framework} 
For the considered decay, the related weak effective Hamiltonian can be written as 
\begin{equation}
\begin{array}{rcl}
H_{eff}=\frac{G_{F}}{\sqrt{2}}V_{cb}V^{*}_{ud}(C_{1}(\mu) O_{1}+C_{2}(\mu)O_{2}),
\end{array}
\end{equation}
with the Fermi constant $ G_{F} $, Cabibbo-Kobayashi-Maskawa(CKM)
matrix elements V and Wilson coefficients $ C_{i}(\mu) $ at the
renormalization scale $ \mu $. \\
The local operators are written as
\begin{eqnarray}
O_{1}=(\bar{q}_{i}u_{i})_{V-A}(\bar{c}_{j}b_{j})_{V-A} ,\nonumber\\
O_{2}=(\bar{q}_{i}u_{j})_{V-A}(\bar{c}_{j}b_{i})_{V-A},
\end{eqnarray}
where i and j are color indices. The operators $ O_{1} $
and $ O_{2} $ are called the current-current operators.
The hard part of the pQCD approach contains six quarks rather than four quarks. This 
is called six-quark effective theory or six-quark operator.A $ B_{c}\rightarrow M_{2}M_{3} $ 
decay amplitude is factorized into the convolution of the six-quark hard
kernel (H), the jet function (J) and the Sudakov factor (S) with the bound-state wave
functions $ (\Phi) $ as follows,
\begin{eqnarray}
A(B_{c}\rightarrow M_{2}M_{3}) &=& \Phi_{B_{c}}\otimes H \otimes J \otimes S \otimes \Phi_{M_{2}} \otimes \Phi_{M_{3}}.
\end{eqnarray}
In the practical applications to heavy $ B_{c} $ meson decays, the decay amplitude of Eq. (3)
in the pQCD approach can be written as,
\begin{eqnarray}
A(B_{c}\rightarrow M_{2}M_{3} ) &\sim & \int d^{4}k_{1}d^{4}k_{2}d^{4}k_{3}Tr[\Phi_{B_{c}}(k_{1})\Phi_{M_{2}}(k_{2})\nonumber \\&& \times
\Phi_{M_{3}}(k_{3})H(k_{1},k_{2},k_{3},t)], \ \ 
\end{eqnarray}
where $ k_{3}’s $ are momenta of light quarks included in each mesons, and Tr denotes the trace
over Dirac and color indices. The function $ H(k_{1},k_{2},k_{3},t) $ describes the four quark operator and the spectator
quark connected by a hard gluon whose $ q_{2} $ is in the order of $ \lambda_{QCD}m_{Bc} $. The wave function $\Phi_{B_{c}}(k_{1}) $ and $\Phi_{M_{i}} $describe
the hadronization of the quark and anti-quark in the $ B_{c} $ meson and the final state light
meson $ M_{i} $.\\
The two final state meson momenta can be written as
\begin{eqnarray}
P_{1}=\frac{m_{B_{c}}}{\sqrt{2}}(1,1,0_{T}), \ \ \ P_{2}=\frac{m_{B_{c}}}{\sqrt{2}}(1-r_{3}^{2},r_{2}^{2},0_{T}), \ \ \ P_{2}=\frac{m_{B_{c}}}{\sqrt{2}}(r_{3}^{2},1-r_{2}^{2},0_{T}),
\end{eqnarray}
where $ r_{i}={m_{i}}/{M_{B_{c}}} $. Putting the quark momenta in Bq, $ M_{2} $ and $ M_{3} $ meson as $ K_{1} $, $ K_{2} $, and $ K_{3} $, respectively, we can choose
\begin{eqnarray}
k_{1}=(x_{1}P^{+}_{1},0, k_{1T}), \ \ \ k_{2}=(x_{2}P^{+}_{2},0, k_{2T}), \ \ \ k_{3}=(0,x_{3}P^{-}_{3}, k_{3T}).
\end{eqnarray}
Then the integration over $ k_{1}^{-} $ , $ k_{2}^{-} $ , and $ k_{3}^{+} $ will 
lead to the decay amplitude in the pQCD approach,
\begin{eqnarray}
A(B_{c}\rightarrow M_{1} M_{2}) &\sim & \int dx_{1}dx_{2}dx_{3} b_{1}db_{1}b_{2}db_{2}b_{3}db_{3} \nonumber \\ && \times Tr[C(t)\Phi_{B_{c}}(x_{1},b_{1})\Phi_{M_{2}}(x_{2},b_{2})\Phi_{M_{3}}(x_{3},b_{3})]
\nonumber \\ && \times H(x_{i},b_{i},t)S_{t}(x_{i}e^{-S(t)}],
\end{eqnarray}
where $ b_{i} $ is the conjugate space coordinate of $ k_{iT} $, and t is the largest energy scale in function
$ H(x_{i},b_{i},t) $ . The large logarithms $ ln(m_{W}/t) $ are included in the Wilson coefficients
$ C(t) $. The large double logarithms $ (ln^{2}x_{i}) $ are summed by the threshold resummation,
and they lead to the jet function $ S_{t}(x_{i}) $ which smears the end-point singularities on $x_{i}$~\cite{x}. The
last term, $ e^{-S(t)} $, is the Sudakov factor which suppresses the soft dynamics effectively~\cite{t3}.\\ 
The longitudinal and transverse polarization vectors of the axial-vector meson are defined by:
\begin{eqnarray}
\epsilon^{*(0)\mu}&=&\frac{E}{m_{V(A)}}[(1-\frac{m^{2}_{V(A)}}{4E^{2}})n_{-}^{\mu}-\frac{m_{V}^{2}}{4E^{2}}n_{+}^{\mu}],\nonumber\\
\epsilon^{*(\lambda)\mu}_{\perp}&=&(\epsilon^{*(\lambda)\mu}-\frac{\epsilon^{*(\lambda)}n_{+}}{2}n_{-}^{\mu}-\frac{\epsilon^{*(\lambda)}n_{-}}{2}n_{+}^{\mu})\delta_{\lambda,\pm 1},
\end{eqnarray}
where $ n_{-}^{\mu}\equiv (1,0,0,-1) $ and $ n_{+}^{\mu}\equiv (1,0,0,1) $,
in the
$ B $ rest frame, for a axial-vector meson is moving along the z-axis, while the $ x $ axes of both daughter particles are parallel the coordinate systems in the Jackson convention~\cite{t}:
\begin{equation}
\begin{array}{rclcrcl}
\epsilon_{1}^{\mu(0)}&=&(p_{c},0,0,E_{1})/m_{1} , && \epsilon_{2}^{\mu(0)}&=&(p_{c},0,0,-E_{2})/m_{2},
\nonumber\\
\epsilon_{1}^{\mu(\pm 1)}&=&\frac{1}{\sqrt{2}}(0,\pm 1,-i,0), && \epsilon_{2}^{\mu(\pm 1)}&=&\frac{1}{\sqrt{2}}(0,\mp 1,i,0),
\nonumber\\
\end{array}
\end{equation}
where $ p_{c} $ is the center mass momentum of the final state meson and $ \epsilon^{*(\pm1)}_{1}.\epsilon^{*(\pm1)}_{2}=-\delta _{\pm1},\delta _{\pm1} $. If the A meson moves along the $ n_{-}^{\mu} $ In the large energy limit, we have $ \epsilon_{A}^{*(\lambda)}.n_{+}=2E_{A}/m_{A}\delta _{\lambda,0} $ and $ \epsilon_{A}^{*(\lambda)}.n_{-}=0 $. If the coordinate systems are in the Jacob-Wick convention where
the y axes of both decay particles are parallel, the transverse polarization vectors of the second
meson will become $ \epsilon_{2}^{\mu}=(0,\pm1,-i,0)/\sqrt{2} $ and $ \epsilon^{*(\pm1)}_{1}.\epsilon^{*(\pm1)}_{2}=\delta _{\pm1},\delta _{\pm1} $.\\
For the wave function of $ {B_{c}} $ meson, we adopt the form (see Ref~\cite{qw}. and
references therein)as, 
\begin{equation}
\Phi_{B_{c}}=\frac{i}{\sqrt{2N_{c}}}[(\not {P}+m_{B_{c}})\gamma_{5}\phi_{B_{c}}(x)]_{\alpha\beta}.
\end{equation}
where $ P$ is the momentum of the $ B_{c} $ meson and x denotes the momentum fraction
of the c quark in the $ B_{c} $ meson.
The distribution amplitude $ B_{c} $ would be close to $ \delta(x-\frac{m_{c}}{m_{B_{c}}}) $ in the nonrelativistic limit because 
$ B_{c} $ meson embraces b and c
quarks simultaneously.
\\
We therefore adopt the
non-relativistic approximation form for $ \phi_{B_{c}}$ as~\cite{qw1,bc},
\begin{eqnarray}
\phi_{B_{c}}=\frac{f_{B_{c}}}{2\sqrt{2N_{c}}}\delta(x-\frac{m_{c}}{m_{B_{c}}}),
\end{eqnarray}
here b is the conjugate space coordinate of
transverse momentum $ k_{T} $, $ f_{B_{c}} $ and $ N_{c}=3 $ are the decay constant of $ B_{c} $ meson and the color number respectively.\\
wave function $ \phi_{B_{c}} $ is normalized by decay constant of $ f_{B_{c}} $
\begin{eqnarray}
\int_{0}^{1}dx\phi_{B_{c}}(x,b=0)=\frac{f_{B_{c}}}{2\sqrt{2N_{c}}},
\end{eqnarray}
For the vector $ J/\psi $ meson(V), we take the wave function as follows,
\begin{eqnarray}
\Phi^{L}_{J/\psi}(x)&=&\frac{1}{\sqrt{2N_{c}}}\{m_{J/\psi}\not{\epsilon}_{L}\phi^{L}(x)+\not{\epsilon}_{L} \not{p} \phi^{t}(x)\},
\nonumber\\
\Phi^{T}_{J/\psi}(x)&=&\frac{1}{\sqrt{2N_{c}}}\{m_{J/\psi}\not{\epsilon}_{L}\phi^{V}(x)+\not{\epsilon}_{T} \not{p} \phi^{T}(x)\}.
\end{eqnarray}
The asymptotic distribution amplitudes of $ J/\psi $ meson read as~\cite{qw2}:
\begin{eqnarray}
\phi^{L}(x)&=&\phi^{T}(x)=9.58\frac{f_{j/\psi}}{2\sqrt{2N_{c}}}x(1-x)[\frac{x(1-x)}{1-2.8x(1-x)}]^{0.7},
\nonumber\\
\phi^{t}(x)&=&10.94\frac{f_{j/\psi}}{2\sqrt{2N_{c}}}(1-2x)^{2}[\frac{x(1-x)}{1-2.8x(1-x)}]^{0.7},
\nonumber\\
\phi^{V}(x)&=&1.67\frac{f_{j/\psi}}{2\sqrt{2N_{c}}}[1+(1+2x)^{2}][\frac{x(1-x)}{1-2.8x(1-x)}]^{0.7}.
\end{eqnarray}
Here, $ \phi^{L} $ and $\phi^{T} $ denote for the twist-2 distribution amplitudes, and $ \phi^{t} $ and $ \phi^{V} $ for the twist-3. x denotes the momentum fraction of the charm quark inside the
charmonium. In which the twist-3 ones $\phi^{t,V} $ vanish, as the twist-2 ones, at the end points due to the factor $ [x(1-x)]^{0.7} $. \\
The twist-2 distribution amplitudes for the longitudinally and trasversely polarized
axial-vector mesons (A) can be parameterized as\cite{o,a2}:
\begin{eqnarray}
\phi_{A}(x)&=&\frac{3f_{A}}{\sqrt{2N_{c}}}x(1-x)\{a_{0A}^{\parallel}+3a_{1A}^{\parallel}(2x-1)+a_{2A}^{\parallel}\frac{3}{2}(5(2x-1)^{2}-1)\}\nonumber\\
\phi_{A}^{T}(x)&=&\frac{3f_{A}}{\sqrt{2N_{c}}}x(1-x)\{a_{0A}^{\perp}+3a_{1A}^{\perp}(2x-1)+a_{2A}^{\perp}\frac{3}{2}(5(2x-1)^{2}-1)\}.
\end{eqnarray}
These distribution amplitudes
satisfy the relation
\begin{eqnarray}
\int_{0}^{1}\phi_{^{3}P_{1}}(x) =\frac{f_{}^{3}P_{1}}{2\sqrt{2N_{c}}},\nonumber\\
%\int_{0}^{1}\phi_{^{3}P_{1}}(x) =a_{0}^{\perp}_{^{3}P_{1}}\frac{f_{}^{3}P_{1}}{2\sqrt{2N_{c}}},\nonumber\\
%\int_{0}^{1}\phi_{^{1}P_{1}}(x) =a_{0}^{\parallel}_{^{1}P_{1}}\frac{f_{}^{1}P_{1}}{2\sqrt{2N_{c}}},\nonumber\\
\int_{0}^{1}\phi_{^{1}P_{1}}(x) =\frac{f_{}^{1}P_{1}}{2\sqrt{2N_{c}}}.
\end{eqnarray}
As for twist-3 distribution amplitudes for axial-vector meson, we use the following form
\begin{eqnarray}
\phi^{t}_{A}(x)&=&\frac{3f_{A}}{2\sqrt{2N_{c}}}\{a_{0A}^{\perp}(2x-1)^{2}+\frac{1}{2}a_{1A}^{\perp}(2x-1)(3(2x-1)^{2}-1)\},\nonumber\\
\phi^{s}_{A}(x)&=&\frac{3f_{A}}{2\sqrt{2N_{c}}}\frac{d}{dx}\{x(1-x)(a_{0A}^{\perp}+a_{1A}^{\perp}0))\},\nonumber\\
\phi^{v}_{A}(x)&=& \frac{3f_{A}}{4\sqrt{2N_{c}}}\{\frac{1}{2}a_{0A}^{\parallel}(1+(2x-1)^{2})
+a_{1A}^{\parallel}(2x-1)^{3}\},\nonumber\\
\phi^{a}_{A}(x)&=&\frac{3f_{A}}{4\sqrt{2N_{c}}}\frac{d}{dx}\{x(1-x)(a_{0A}^{\parallel}+a_{1A}^{\parallel}(2x-1))\}.
\end{eqnarray}
Here x represents the momentum fraction carried by quark in the meson and $ a_{0}^{\parallel}{}_{{3}P_{1}}=1 $, $ a_{0}^{\perp}{}_{{3}P_{1}}=0 $. \\
The intrinsic $b$ dependence for the heavy meson wave
function $ \phi_{B_{c}} $ is important and for the light axial-vector meson wave function $\phi_{A}$ is negligible\cite{o1}.
\section{Branching ratio of $ B^{+}_{c}\rightarrow J/\psi \ a_{1}(1260)^{+}$ decay }
In the pQCD approach, the four Feynman diagrams for $ B^{+}_{c}\rightarrow J/\psi \ a_{1}(1260)^{+}$ decays are shown in Fig.1. There are three kinds of polarizations of a axial-vector meson, namely, longitudinal 
$ (L) $, normal $ (N) $, and transverse $ (T) $.
The decay amplitudes for the factorizable diagram (a) and (b) can be read as,\\
(i) $(V-A)(V-A) $
\begin{eqnarray}
{F}^{L}_{fs}&=&8\pi C_{F} f_{A}m^{2}_{B_{c}}\int_{0}^{1}dx_{1}dx_{3}\int_{0}^{\infty}b_{1}db_{1}b_{3}db_{3}\phi_{B_{c}}(x_{1},b_{1})\nonumber\\
&&\times \{[(1+x_{3})\phi^{V}(x_{3})+r_{V}(1-2x_{3})(\phi^{L}(x_{3})+\phi^{t}(x_{3}))]\nonumber\\
&&\times E_{fs}(t_{a})h_{fs}(x_{1},x_{3},b_{1},b_{3})+ 2r_{V}\phi^{T}(x_{3}))\nonumber\\
&&\times E_{fs}(t_{b})h_{fs}(x_{3},x_{1},b_{3},b_{1}) \},
\end{eqnarray}
where $C_{F}=4/3$ is a color factor, $r_{i}={m_i}/{m_{B}}$, and $ f_{A}$ is the decay constant of $ a_{1} $ meson. The convolution functions $E_{ef}$, the hard function $h_{ef}$ and the factorization hard scal $t_{a,b}$ are given in the appendix.\\
(ii) $(V-A)(V+A)$
\begin{equation}
\begin{array}{rcl}
{F}^{L;P_{1}}_{fs}&=&-{F}^{L}_{fs},
\end{array}
\end{equation}
which is originated from $ \langle a_{1}\vert V+A \vert0\rangle =- \langle a_{1}\vert V-A \vert0\rangle $.\\
(iii) $(S-P)(S+P)$
\begin{eqnarray}
{F}^{L;P_{2}}_{fs}&=0,
\end{eqnarray}
because the emitted axial-vector meson can not be produced through a scalar or a pseudoscalar current.\\
For the non-factorizable diagrams (c) and (d), the decay amplitude can be read as \\
(i) $(V-A)(V-A)$
\begin{eqnarray}
{M}^{L}_{nfs}&=&\frac{32}{\sqrt{6}}\pi C_{F} m^{2}_{B_{c}}\int_{0}^{1}dx_{1}dx_{2}dx_{3}\int_{0}^{\infty}
b_{1}db_{1}b_{2}db_{2}\phi_{B_{c}}(x_{1},b_{1})\nonumber\\ && \times \phi_{A}(x_{2})\{[(1-x_{2}\phi^{V}(x_{3})-r_{V}x_{3}(\phi^{L}(x_{3})-\phi^{t}(x_{3}))]\nonumber\\ && \times E_{nfs}(t_{c})h_{nfs}(x_{1}, 1-x_{2},x_{3},b_{1},b_{2})- [(x_{2}+x_{3})\phi^{V}(x_{3})\nonumber\\ && -r_{V}x_{3}(\phi^{V}(x_{3})-r_{V}x_{3}(\phi^{T}(x_{3})+\phi^{t}(x_{3}))])\nonumber\\&& \times E_{nfs}(t_{d}) h_{nfs}(x_{1},x_{2},x_{3},b_{1},b_{2})\},
\end{eqnarray}
(ii) $(V-A)(V+A)$
\begin{eqnarray}
{M}^{L;P_{1}}_{nfs}&=&\frac{32}{\sqrt{6}}\pi C_{F} m^{2}_{B_{c}}\int_{0}^{1}dx_{1}dx_{2}dx_{3}\int_{0}^{\infty}
b_{1}db_{1}b_{2}db_{2}\phi_{B_{c}}(x_{1},b_{1})\nonumber\\
&&\times r_{V}\{[(1-x_{2})(\phi_{A}^{s}(x_{2})+\phi_{A}^{t}(x_{2}))\phi^{V}(x_{3})-r_{V}(\phi_{A}^{s}(x_{2})\nonumber\\
&&\times[(x_{2}-x_{3}-1)\phi^{L}(x_{3})-(x_{2}+x_{3}-1)\phi^{t}(x_{3})]+\phi_{A}^{t}(x_{2})\nonumber\\
&&\times[(x_{2}+x_{3}-1)\phi^{T}(x_{3})+(1-x_{2}+x_{3})\phi^{t}(x_{3}])]\nonumber\\
&&\times E_{nfs}(t_{c})h_{nfs}(x_{1},1-x_{2},x_{3},b_{1},b_{2})-
[x_{2}(\phi_{A}^{s}(x_{2})\nonumber\\ && -\phi_{A}^{t}(x_{2}))\phi^{V} +r_{V}(x_{2}(\phi_{A}^{s}(x_{2})-\phi_{A}^{t}(x_{2}))(\phi^{L}(x_{3})-\phi^{t}(x_{3}))
\nonumber\\ &&+x_{3}(\phi_{A}^{s}(x_{2})+\phi_{A}^{t}(x_{2}))(\phi^{V}(x_{3})+\phi^{t}(x_{3})))]\nonumber\\ &&\times E_{nfs}(t_{d})h_{nfs} (x_{1},x_{2},x_{3},b_{1},b_{2})\},
\end{eqnarray}
%\begin{figure}
%\centering
%\includegraphics[scale=.78]{bbbbb}
%\caption{Diagrams contributing to the decay $ B^{+}_{c}\longrightarrow J/\psi \ a_{1}(1260)^{+}$. }
%\label{f1}
%\end{figure}
(iii) $(S-P)(S+P)$
\begin{eqnarray}
{M}^{L;P_{2}}_{nfs}&=&\frac{32}{\sqrt{6}}\pi C_{F} m^{2}_{B_{c}}\int_{0}^{1}dx_{1}dx_{2}dx_{3}\int_{0}^{\infty}
b_{1}db_{1}b_{2}db_{2}\phi_{B_{c}}(x_{1},b_{1})\nonumber \\
&&\times \phi_{A}(x_{2})\{[(x_{2}-x_{3}-1)\phi^{V}(x_{3})+r_{V}x_{3}(\phi^{L}(x_{3})\nonumber \\
&&+\phi^{t}(x_{3}))] E_{nfs}(t_{c}) h_{enf}(x_{1}, 1-x_{2},x_{3},b_{1},b_{2})\nonumber\\ &&+
[x_{2}\phi^{V}(x_{3})-r_{V}x_{3}(\phi^{L}(x_{3})-\phi^{t}(x_{3}))]\nonumber\\ && \times E_{nfs}(t_{d}) h_{enf}(x_{1},x_{2},x_{3},b_{1},b_{2})\},
\end{eqnarray}
We can also present the factorization formulas for the Feynman amplitudes with trans-
verse polarizations,
\begin{eqnarray}
{F}^{N}_{fs}&=&8\pi C_{F} f_{A}m^{2}_{B_{c}}\int_{0}^{1}dx_{1}dx_{3}\int_{0}^{\infty}b_{1}db_{1}b_{3}db_{3}\phi_{B_{c}}(x_{1},b_{1})r_{V}\nonumber\\
&&\times \{[\phi^{t}(x_{3})+r_{V}x_{3}(\phi^{V}(x_{3})-\phi^{L}(x_{3}))+2r_{V}\phi^{V}(x_{3})]\nonumber\\
&&\times E_{fs}(t_{a})h_{fs}(x_{1},x_{3},b_{1},b_{3})+r_{V}(\phi^{T}(x_{3})+\phi^{V}(x_{3})\nonumber\\
&&\times E_{fs}(t_{b}) h_{fs}(x_{3},x_{1},b_{3},b_{1})\},
\end{eqnarray}
\begin{eqnarray}
{F}^{T}_{fs}&=&16\pi C_{F} f_{A}m^{2}_{B_{c}}\int_{0}^{1}dx_{1}dx_{3}\int_{0}^{\infty}b_{1}db_{1}b_{3}db_{3}\phi_{B_{c}}(x_{1},b_{1})r_{V}\nonumber\\
&&\times \{[\phi^{t}(x_{3})-r_{V}x_{3}(\phi^{V}(x_{3})-\phi^{L}(x_{3}))+2r_{V}\phi^{T}(x_{3})]\nonumber\\
&&\times E_{fs}(t_{a})h_{fs}(x_{1},x_{3},b_{1},b_{3})+r_{V}(\phi^{L}(x_{3})+\phi^{V}(x_{3})\quad \quad\quad \quad \quad\nonumber\\
&&\times E_{fs}(t_{b})h_{fs}(x_{3},x_{1},b_{3},b_{1})\},
\end{eqnarray}
\begin{equation}
\begin{array}{rcl}
{F}^{N;P_{1}}_{fs}&=&-{F}^{N}_{fs},\nonumber\\
{F}^{T;P_{1}}_{fs}&=&-{F}^{T}_{fs},\nonumber\\
\\
{F}^{N;P_{2}}_{fs}&=&0,\nonumber\\
{F}^{T;P_{1}}_{fs}&=&0,\nonumber\\
\end{array}
\end{equation}
\begin{eqnarray}
{M}^{N}_{nfs}&=&\frac{32}{\sqrt{6}}\pi C_{F} m^{2}_{B_{c}}\int_{0}^{1}dx_{1}dx_{2}dx_{3}\int_{0}^{\infty}b_{1}db_{1}b_{2}db_{2}\phi_{B_{c}}(x_{1},b_{1})r_{A}\nonumber\\
&&\times \{[(1-x_{2})(\phi_{A}^{a}(x_{2})+\phi_{A}^{v}(x_{2}))\phi^{t}(x_{3})] h_{nfs}(x_{1},1-x_{2},x_{3},b_{1},b_{2})\nonumber\\
&&\times E_{nfs}(t_{c})+ 
[x_{2}(\phi_{A}^{a}(x_{2})+\phi_{A}^{v}(x_{2}))\phi^{t}(x_{3}) -2r_{V}(x_{2}+x_{3})
\nonumber\\
&&\times (\phi_{A}^{a}(x_{2})\phi_{V}^{a}(x_{3})+\phi_{A}^{v}(x_{2}))]E_{nfs}(t_{d})h_{nfs}(x_{1},x_{2},x_{3},b_{1},b_{2})\},
\end{eqnarray}
\begin{eqnarray}
{M}^{T}_{nfs}&=&\frac{64}{\sqrt{6}}\pi C_{F} m^{2}_{B_{c}}\int_{0}^{1}dx_{1}dx_{2}dx_{3}\int_{0}^{\infty}b_{1}db_{1}b_{2}db_{2}\phi_{B_{c}}(x_{1},b_{1})r_{A}\nonumber\\
&&\times \{[(1-x_{2})(\phi_{A}^{a}(x_{2})+\phi_{A}^{v}(x_{2}))\phi^{t}(x_{3})]E_{nfs}(t_{c})\nonumber\\
&&\times h_{nfs}(x_{1}, 1-x_{2},x_{3},b_{1},b_{2}) +[x_{2}(\phi_{A}^{a}(x_{2}) +\phi_{A}^{v}(x_{2}))\nonumber\\ && \times
\phi^{t}(x_{3}) -2r_{V}(x_{2}+x_{3}) (\phi_{A}^{a}(x_{2}) \phi^{L}(x_{3})
+\phi_{A}^{v}(x_{2}))]\quad \quad\quad \quad \quad \nonumber\\ && \times E_{nfs}(t_{d})h_{nfs}(x_{1},x_{2},x_{3},b_{1},b_{2})\},
\end{eqnarray}
\begin{eqnarray}
{M}^{N;P_{1}}_{nfs}&=&-\frac{32}{\sqrt{6}}\pi C_{F} m^{2}_{B_{c}}\int_{0}^{1}dx_{1}dx_{2}dx_{3}\int_{0}^{\infty}b_{1}db_{1}b_{2}db_{2}\phi_{B_{c}}(x_{1},b_{1})r_{A}\nonumber\\
&&\times x_{3}\phi_{A}^{T}(x_{2})(\phi^{L}(x_{3})-\phi^{V}(x_{3}))[ E_{nfs}(t_{c})h_{nfs}(x_{1}, 1-x_{2},x_{3},b_{1},b_{2})\nonumber\\
&&+E_{nfs}(t_{d})h_{nfs}(x_{1},x_{2},x_{3},b_{1},b_{2})],
\end{eqnarray}
\begin{eqnarray}
{M}^{T;P_{1}}_{nfs}&=&-\frac{64}{\sqrt{6}}\pi C_{F} m^{2}_{B_{c}}\int_{0}^{1}dx_{1}dx_{2}dx_{3}\int_{0}^{\infty}b_{1}db_{1}b_{2}db_{2}\phi_{B_{c}}(x_{1},b_{1})r_{A}\nonumber\\
&&\times x_{3}\phi_{A}^{T}(x_{2})(\phi^{L}(x_{3})-\phi^{V}(x_{3}))[E_{nfs}(t_{c})h_{nfs}(x_{1}, 1-x_{2},x_{3},b_{1},b_{2})\nonumber\\
&&+E_{nfs}(t_{d})h_{nfs}(x_{1},x_{2},x_{3},b_{1},b_{2})],
\end{eqnarray}
\begin{eqnarray}
{M}^{N;P_{2}}_{nfs}&=&-\frac{32}{\sqrt{6}}\pi C_{F} m^{2}_{B_{c}}\int_{0}^{1}dx_{1}dx_{2}dx_{3}\int_{0}^{\infty}b_{1}db_{1}b_{2}db_{2}\phi_{B_{c}}(x_{1},b_{1})r_{A}\nonumber\\
&&\times \{[x_{2}(\phi_{A}^{a}(x_{2})-\phi_{A}^{v}(x_{2}))\phi^{t}(x_{3})]
E_{nfs}(t_{d})h_{nfs}(x_{1}, 1-x_{2},x_{3},b_{1},b_{2})\nonumber\\&&\times 
[x_{2}(\phi_{A}^{a}(x_{2})-\phi_{A}^{v}(x_{2}))\phi^{t}(x_{3})+2r_{V}(1-x_{2}+x_{3})(\phi_{A}^{v}(x_{2})\phi^{V}(x_{3})\nonumber\\
&&-
\phi_{A}^{a}(x_{2})\phi^{L}(x_{3}))]E_{nfs}(t_{c})h_{nfs}(x_{1},x_{2},x_{3},b_{1},b_{2})\},
\end{eqnarray}
\begin{eqnarray}
{M}^{T;P_{2}}_{nfs}&=&-\frac{64}{\sqrt{6}}\pi C_{F} m^{2}_{B_{c}}\int_{0}^{1}dx_{1}dx_{2}dx_{3}\int_{0}^{\infty}b_{1}db_{1}b_{2}db_{2}\phi_{B_{c}}(x_{1},b_{1})r_{A}\nonumber\\
&&\times \{[x_{2}(\phi_{A}^{a}(x_{2})-\phi_{A}^{v}(x_{2}))\phi^{t}(x_{3})]h_{nfs}(x_{1}, 1-x_{2},x_{3},b_{1},b_{2})\nonumber\\
&&\times E_{nfs}(t_{d})+[x_{2}(\phi_{A}^{a}(x_{2})-\phi_{A}^{v}(x_{2}))\phi^{t}(x_{3})
+2r_{V}(1-x_{2}+x_{3})\nonumber\\&& \times(\phi_{A}^{v}(x_{2})\phi^{V}(x_{3})
\phi_{A}^{a}(x_{2})\phi^{L}(x_{3}))]h_{nfs}(x_{1},x_{2},x_{3},b_{1},b_{2})E_{nfs}(t_{c})\}.
\end{eqnarray}
Decay amplitudes with three polarizations $ h=L, N, T $ as follows,
\begin{eqnarray}
M&=&V_{cb}V_{ud}^{*}[(C_{1}+\frac{C_{2}}{3})F_{fs}^{h}+C_{2}M_{nfs}^{h}],
\end{eqnarray}
the decay rate can be written explicitly as
\begin{equation}
\Gamma= \frac{G_{F}^{2}P}{16\pi m_{B_{c}}^{2}}\sum_{\sigma=L, N, T}M^{\sigma \dagger}M^{\sigma},
\end{equation}
where $ G_{F}$ is the Fermi coupling constant. $ V_{cb}$ and $ V_{ud}$ are the Cabibbo-
Kobayashi-Maskawa (CKM) matrix factors and $C_{i} $ are the Wilson coefficients.\\
The branching ratio is given by
\begin{equation}
BR=\Gamma \times \tau_{B_{c}}.
\end{equation}
The decay amplitudes $ M^{\sigma} $ can be described by
\begin{eqnarray}
M^{\sigma}=\epsilon_{2\mu}^{\ast}(\sigma)\epsilon_{3\nu}^{\ast}(\sigma)[ag^{\mu\nu}+\frac{b}{m_{2}m_{3}}P_{B_{c}}^{\mu}P_{B_{c}}^{\nu}+i\frac{c}{m_{2}m_{3}}\epsilon^{\mu\nu\alpha\beta}P_{2}^{\alpha}P_{3}^{\beta}]\quad\quad\quad\quad\nonumber\\
=M_{L}+M_{N}\epsilon_{2}^{\ast}(\sigma=T).\epsilon_{3}^{\ast}(\sigma=T)+i\frac{M_{T}}{m_{B_{c}}^{2}}\epsilon^{\alpha\beta\gamma\rho}\epsilon_{2\alpha}^{\ast}(\sigma)\epsilon_{2\beta}^{\ast}(\sigma)P_{2\gamma}
P_{3\rho},
\end{eqnarray}
the subscript $ \sigma $ denotes
the helicity states of the two final mesons.
The amplitude $ M^{\sigma} $ can be decomposed, according to the Lorentz-invariant amplitudes a, b and c \cite{am}
\begin{eqnarray}
m_{B_{c}}^{2}M_{L}=a\epsilon_{2}^{\ast}(L).\epsilon_{3}^{\ast}(L)+\frac{b}{m_{2}m_{3}}\epsilon_{2}^{\ast}(L).P_{3}\epsilon_{2}^{\ast}(L).P_{2},
\nonumber\\m_{B_{c}}^{2}M_{N}=a,\quad\quad\quad\quad\quad\quad\quad\quad\quad\quad\quad\quad\quad\quad\quad\nonumber\\
m_{B_{c}}^{2}M_{T}=\frac{m_{B_{c}}^{2}}{m_{2}m_{3}}c.\quad\quad\quad\quad\quad\quad\quad\quad\quad\quad\quad\quad\quad
\end{eqnarray}
We can define the amplitudes $ A_{(i=L,\parallel,\perp)} $ as
\begin{eqnarray}
A_{L}=-\xi m_{B_{c}}^{2}M_{L},\nonumber\\
A_{\parallel}=\xi\sqrt{2}m_{B_{c}}^{2}M_{N},\nonumber\\
A_{\perp}=\xi r_{2}r_{3}\sqrt{2(r^{2}-1)}m_{B_{c}}^{2}M_{T},
\end{eqnarray}
for the longitudinal, parallel, and perpendicular polarizations, respectively, with the normalization
factor $\xi=\sqrt{ G_{F}^{2}p_{c}/ (16\pi m_{B_{c}}^{2}\Gamma)} $ and $ r=P_{2}P_{3}/(m_{2}m_{3}) $.\\
These amplitudes satisfy the relation,
\begin{eqnarray}
\vert A_{L} \vert ^{2}+\vert A_{\parallel} \vert ^{2}+\vert A_{\perp} \vert ^{2}=1.
\end{eqnarray}
The polarization fractions $f_{L} $, $ f_{\parallel} $ and $f _{\perp} $ canbe defined as
\begin{eqnarray}
f_{L(\parallel,\perp)} =\frac{\vert A_{L(\parallel,\perp) } \vert^{2}}{\vert A_{L} \vert ^{2}+\vert A_{\parallel} \vert ^{2}+\vert A_{\perp} \vert ^{2}}=\vert A_{L(\parallel,\perp) } \vert^{2}.
\end{eqnarray}
The direct CP asymmetry $ A_{cp}^{dir} $ is defined as\cite{1000}
\begin{eqnarray}
A_{cp}^{dir,\alpha}=\frac{\bar{f_{\alpha}}-f_{\alpha}}{\bar{f\alpha}+f_{\alpha}}(\alpha=L,\parallel,\perp).
\end{eqnarray}
\section{Numerical results }
We use other input parameters as follows:\\ 
$m_{B_{c}}=6.277 \ GeV $, $ m_{a_{1}}=1.23 \ GeV$, $m_j/\psi=3.096\ GeV$, $ m_{c}=1.27\ GeV $, $ m_{W}=80.4\ GeV $, 
$f_{B_{c}}=(0.489\pm0.004) \ GeV$, $f_{j/\psi}=(0.405 \pm0.014) \ GeV$, $f_{a_{1}}=0.238 \ GeV$, $a^{\parallel, a_{1}}_{2}=-0.02\pm0.02 $, $ a^{\perp, a_{1}}_{1}=-1.04\pm0.34 $, 
$\tau_{B_{c}}=0.46\ ps$, $\Lambda_{QCD} =0.2250 \ GeV$, $ G_{F}=1.16\times 10^{-5} $ \ ~\cite{e,EXP,api,oo}
\\
The CKM matrix is a $ 3\times3 $ unitary matrix as~\cite{EXP} \\
\begin{eqnarray}
V=
\begin{pmatrix}
V_{ud} & V_{us} & V_{ub} \\ 
V_{cd} & V_{cs} & V_{cb} \\ 
V_{td} & V_{ts} & V_{tb} \nonumber\\
\end{pmatrix}
\end{eqnarray}
The elements of the CKM matrix can be parameterized by three mixing angles $ A, \lambda, \rho $ and a
CP-violating phase $ \eta $
\begin{eqnarray}
V=
\begin{pmatrix}
1-\lambda^{2}/2 & \lambda & A\lambda^{3}(\rho-i\eta) \\ 
-\lambda & 1-\lambda^{2}/2& A\lambda^{2} \\ 
\lambda^{3}(1-\rho-i\eta) & -A\lambda^{2} & 1\nonumber\\
\end{pmatrix}
\end{eqnarray}
The results for the Wolfenstein parameters are
\begin{eqnarray}
\lambda=0.22535 \pm 0.00065,\nonumber A = 0.811^{+0.022} _{-0.012},
\nonumber\\
\bar{\rho} =0.131^{+0.026} _{-0.013},\quad\quad\quad\quad\quad\quad\quad \bar{\eta} =0.345^{+0.013} _{-0.014}.\nonumber
\end{eqnarray}
We use the central values of the Wolfenstein parameters and obtain
\begin{eqnarray}
V_{cb}=0.0412 ^{+0.0011} _{-0.0005}, \quad\quad\quad\quad V_{ud}=0.97427\pm 0.00015.\nonumber
\end{eqnarray}
%The Wilson coefficients $C_i(\mu)$ have been calculated in different schemes~\cite{fff}. 
The Wilson coefficients $ C_{1} $ and $ C_{2} $, the coupling constants for the interaction terms
$ Q_{1} $ and $ Q_{2} $, become calculable nontrivial functions of $ \alpha_{s} $, $ M_{W} $ and the renormalization
scale $ \mu $.
Wilson coefficients $C(\mu_{W}) $ at $ \mu_{W}=M_{W} $ are $ C_{1}(\mu_{W}) =0$, $ C_{2}(\mu_{W}) =1 $\cite{fff}.
In the generalized factorization
approach, it is well established that non-factorizable contributions must be present in the matrix
elements in order to cancel the scale $ \mu $ and the renormalization scheme dependence of $ C_{i} (\mu)$.
To solve the issue of scale $ \mu $ dependence, but not the renormalization scheme dependence,
to isolate form the matrix element of four operators the $ $
dependence, and link with the $ \mu $ dependence in the Wilson coefficients $ C_{i} (\mu)$.
In table (1)\cite{444}, we simply present the numerical results of
the leading order (LO) and next-to-leading order (NLO) Wilson coefficients in different scales.
We use the NLO Wilson 
coefficient because NLO contributions in the renormalization
group improved perturbation theory.
%We use the NLO Wilson coefficients obtained in the naive-dimensional regularization scheme
%(NDR) at the energy scale $ \mu=m_{b} $. The values are $ C_{1} =1.082$ and $ C_{2}=-0.185 $\cite{444}.
%In the generalized factorization
%approach, it is well established that non-factorizable contributions must be present in the matrix
%elements in order to cancel the scale μ and the renormalization scheme dependence of $ C_{i} (\mu)$.
\begin{table}[h ]
\centering
\caption{
The values of LO and NLO Wilson coefficients $ C_{i}(\mu) $ for
$ \mu=m_{b} $, $ \mu=m_{b}/2 $, $ \mu=2m_{b}$}
\label{tab:1} 
\begin{tabular}{llll}
\hline\noalign{\smallskip}
& $ \mu= m_{b}/2 $& \quad\quad$\mu= m_{b} $&\quad\quad$ \mu= 2m_{b} $ \\
\hline\noalign{\smallskip}
& $ LO \quad\quad NLO $\quad\quad\quad\quad &$ LO \quad\quad NLO $\quad\quad\quad\quad &$ LO \quad\quad NLO $\\
\hline\noalign{\smallskip}
$ C_{1}$& 1.179 \quad 1.115 &1.179 \quad 1.080 & 1.072 \quad 1.043\\
$ C_{2}$& -0.370\quad -0.280 &-0.255 \quad -0.180 & -0.171 \quad -0.104 \\
\noalign{\smallskip}\hline
\end{tabular}
\end{table}
\\
The branching ratio of $ B^{+}_{c}\rightarrow J/\psi \ a_{1}(1260)^{+}$ has 
been calculated at three scales, e.g. $\mu= m_{b} $, $ \mu= m_{b}/2 $ and $\mu= 2m_{b} $ in table (2).
The experimental data of branching ratio of this decay is less than $ 1.3\times 10^{-3} $~\cite{EXP}, the branching at the scale $\mu= m_{b}$ is nearly 
agreement with the experimental result.\\
\begin{table}[h] 
\centering
\caption{Branching ratios of the $ B^{+}_{c}\rightarrow J/\psi \ a_{1}(1260)^{+}$ decay}
\label{tab:1} 
\begin{tabular}{llll}
\hline\noalign{\smallskip}
energy scale& Branching ratios & \\
\noalign{\smallskip}\hline\noalign{\smallskip}
$ \mu= m_{b}/2$ & $(0.97^{+0.05+0.03+0.02}_{-0.06-0.02-0.01})\times10^{-3}$ \\
$\mu= m_{b} $ & $(1.02^{+0.04+0.03+0.01}_{-0.08-0.05-0.01})\times10^{-3}$ \\
$ \mu= 2m_{b} $ & $(1.3^{+0.06+0.02+0.01}_{-0.07-0.04-0.03})\times10^{-3}$ \\
\noalign{\smallskip}\hline
\end{tabular}
\end{table}
For the theoretical uncertainties in our calculation, we estimated three kinds
of them: The first errors in our
calculations are caused by the the decay constants, the shape parameters in wave function $ B_{c} $, Gegenbauer moments of $ a_{1} $
and the hard energy scale $\mu_{i} $;
The second errors are from the unknown next-to-leading order QCD corrections with respect to $ \alpha_{s} $
and
the power corrections, characterized by the choice of the $ \Lambda_{QCD} $;
The third error are estimated from the uncertainties of the CKM matrix elements.\\
We calculate the direct CP-violating asymmetry in every polarization
and we obtain the results in the pQCD approach as
\begin{equation}
A_{cp}^{dir,L}\approx 0.0,\nonumber\\
A_{cp}^{dir,\parallel}\approx 0.0,\nonumber\\ 
A_{cp}^{dir,\perp}\approx 0.0.\nonumber
\end{equation}
There is only one kind of Cabibbo-Kabayashi-Muskawa (CKM) phase involved
in the decay amplitude (see Eq.(31)), therefore CP violation is absent for every polarization in this decay.
\section{conclusion } 
In this work we have presented a detailed study of two-body $ B_{c} $ decays into final
states involving one vector and one
axial-vector meson (VA), within the framework of pQCD approach.
We have calculated the branching ratio of $ B^{+}_{c}\rightarrow J/\psi \ a_{1}(1260)^{+}$ decay in pQCD approach based on the 
$ k_{T} $
factorization theorem. We calculated not only the factorizable emission diagrams, but also
the nonfactorizable spectator.
The experimental data of branching ratio of this decay is lesser than ($ 1.3\times 10^{-3} $). We obtained the branching ratio of pQCD approache is $((1.02^{+0.04+0.03+0.01}_{-0.08-0.05-0.01}) \times10^{-3})$ at scale $ \mu= m_{b} $,
this branching ratio has an 
agreement with the experimental result. 
Our theoretical predictions in the pQCD approach will
provide an important platform for testing the SM and exploring the helicity
structure of this considered decay. It can also provide more information on
measuring the unitary CKM angles and understanding the decay mechanism.

\appendix

\section{Appendix}

The function $ E_{fs} $ and $ E_{nfs} $
defined as
\begin{eqnarray}
E_{fs}(t)&=&\alpha_{s}(t).exp[-S_{B_{c}}(t)-S_{2}(t)]\nonumber\\
E_{nfs}(t)&=&\alpha_{s}(t).exp[-S_{2}(t)-S_{3}(t)].\nonumber
\end{eqnarray}
The Sudakov exponents are defined as
\begin{eqnarray}
S_{B_{c}}(t)&=&s(x_{1}\frac{m_{B_{c}}}{\sqrt{2}},b_{1})+\frac{5}{3}\int_{1/b_{1}}^{t}\frac{d\bar{\mu}}{\bar{\mu}}\gamma_{q}(\alpha_{s}(\bar{\mu})),\nonumber\\
S_{2}(t)&=&s(x_{2}\frac{m_{B_{c}}}{\sqrt{2}},b_{2})+s((1-x_{2})\frac{m_{B_{c}}}{\sqrt{2}},b_{2})+2\int_{1/b_{2}}^{t}\frac{d\bar{\mu}}{\bar{\mu}}\gamma_{q}(\alpha_{s}(\bar{\mu})),\nonumber\\
S_{3}(t)&=&s(x_{3}\frac{m_{B_{c}}}{\sqrt{2}},b_{3})+s((1-x_{3})\frac{m_{B_{c}}}{\sqrt{2}},b_{3})+2\int_{1/b_{3}}^{t}\frac{d\bar{\mu}}{\bar{\mu}}\gamma_{q}(\alpha_{s}(\bar{\mu})),\nonumber\\
s(Q,b)&=&\int_{1/b}^{Q}\frac{d\bar{\mu}}{\bar{\mu}}[\{\frac{2}{3}(2\gamma_{E}-1-log2)+C_{F}log\frac{d\bar{\mu}}{\bar{\mu}}\}\frac{\alpha_{s}}{\pi}\nonumber\\&&
+\{\frac{67}{9}-\frac{\pi^{2}}{3}+\frac{10}{27}n_{f}+\frac{2}{3}\beta_{0}log\frac{\gamma_{E}}{2}\}
(\frac{\alpha_{s}(\bar{\mu})}{\pi})^{2}log\frac{Q}{\bar{\mu}}]\nonumber
\end{eqnarray}
where the function $s(Q,b) $
are defined in the Appendix A of Ref ~\cite{sq}.\\
Here
\begin{eqnarray}
t_{a}&=&max[\sqrt{x_{3}}m_{B_{c}},1/b_{2},1/b_{3}],\nonumber\\
t_{b}&=&max[\sqrt{x_{2}}m_{B_{c}},1/b_{2},1/b_{3}],\nonumber\\
t_{c}&=&max[\sqrt{x_{1}x_{3}}m_{B_{c}},\sqrt{(x_{1}-x_{2})x_{3}} m_{B_{c}},1/b_{1},1/b_{2}],\nonumber\\
t_{d}&=&max[\sqrt{x_{2}x_{3}}m_{B_{c}},\sqrt{\vert x_{1}-x_{2} \vert x_{3}}m_{B_{c}},1/b_{1},1/b_{2}],\nonumber
\end{eqnarray}
\begin{eqnarray}
h_{fs}(x_{1},x_{2},x_{3},b_{2},b_{3})&=&(\frac{i\pi}{2})^{2}H_{0}^{(1)}(\sqrt{x_{2}x_{3}}m_{B_{c}}b_{2})\nonumber\\&& \times
[(\theta(b_{2}-b_{3}))H_{0}^{(1)}(\sqrt{x_{3}}m_{B_{c}}b_{2})J_{0}(\sqrt{x_{3}}m_{B_{c}}b_{3})\nonumber\\&&
+\theta(b_{3}-b_{2}))H_{0}^{(1)}(\sqrt{x_{1}+x_{2}+x_{3}+x_{3}x_{1}}m_{B_{c}}b_{3})\nonumber\\&& \times J_{0}(\sqrt{x_{3}}m_{B_{c}}b_{2})].S_{t}(x_{3})\nonumber
\end{eqnarray}
The threshold resummation form factor $ S_{t}(x_{i)} $ is adopted from Ref~\cite{eq}.
\begin{eqnarray}
S_{t}(x)&=&\frac{2^{l+2c}\Gamma(3/2+c)}{\sqrt{\pi}\Gamma(1+c)}[x(1-x)]^{c},\nonumber
\end{eqnarray}
with $c = 0.3$ in this work.
\begin{eqnarray}
h_{nfs}(x_{1},x_{2},x_{3},b_{1},b_{2})&=&\frac{i\pi}{2}[(\theta(b_{1}-b_{2})H_{0}^{(1)}(\sqrt{x_{2}x_{3}}m_{B_{c}}b_{1})
J_{0}^{(1)}(\sqrt{x_{2}x_{3}m_{B_{c}}b_{2}})\nonumber\\&& \times
\begin{array}{cc}
\frac{i\pi}{2} H_{0}^{(1)} (\sqrt{\vert F_{j}^{2}\vert }m_{B}b_{1}) & F_{j}^{2}\langle 0 \\ 
K_{0}( F_{j} m_{B}b_{1}) & F_{j}^{2}\langle 0\nonumber\\
\end{array} 
\end{eqnarray}
\begin{eqnarray}
F^{2}_{1}&=&1-(1-x_{3})(1-x_{1}-x_{2}),\nonumber\\
F^{2}_{1}&=&x_{3}(x_{1}-x_{2}).\nonumber
\end{eqnarray}
And
\begin{eqnarray}
H_{0}^{(1)}(z)&=&J_{0}(z)+iK_{0}(z)\nonumber
\end{eqnarray}
\end{document}